\documentclass[superscriptaddress,secnumarabic,
amssymb,amsmath,nobibnotes,aps,prd,showkeys,showpacs,nofootinbib]{revtex4}%
\usepackage{graphicx}
\usepackage{epsf}
\usepackage{bm}
\usepackage{amsmath}
\usepackage{amsfonts}
\usepackage{color,amsxtra}
\usepackage{amssymb}%
\providecommand{\U}[1]{\protect\rule{.1in}{.1in}}
\newcommand{\be}{\begin{equation}}
\newcommand{\ee}{\end{equation}}

\newcommand{\mincir}{\raise
-3.truept\hbox{\rlap{\hbox{$\sim$}}\raise4.truept\hbox{$<$}\ }}
\newcommand{\magcir}{\raise
-3.truept\hbox{\rlap{\hbox{$\sim$}}\raise4.truept\hbox{$>$}\ }}

\begin{document}
\title{Two scalar field cosmology: Conservation laws and exact solutions}
\author{Andronikos Paliathanasis}
\email{paliathanasis@na.infn.it}
\affiliation{Dipartimento di Fisica, Universita' di Napoli Federico II, I-80126 Napoli, Italy}
\affiliation{INFN Sezione di Napoli, Complesso Universitario di Monte S. Angelo, Via
Cinthia, 9, I-80126 Napoli, Italy}
\author{M. Tsamparlis}
\email{mtsampa@phys.uoa.gr}
\affiliation{Faculty of Physics, Department of Astrophysics - Astronomy - Mechanics,
University of Athens, Panepistemiopolis, Athens 157 83, Greece}

\begin{abstract}
We consider the two scalar field cosmology in a FRW spatially flat spacetime
where the scalar fields interact both in the kinetic part and the potential.
We apply the Noether point symmetries in order to define the interaction of
the scalar fields. We use the point symmetries in order to write the field
equations in the normal coordinates and we find that the Lagrangian of the
field equations which admits at least three Noether point symmetries describes
linear Newtonian systems. Furthermore, by using the corresponding conservation
laws we find exact solutions of the field equations. Finally, we generalize
our results to the case of N scalar fields interacting both in their potential
and their kinematic part in a flat FRW background.

\end{abstract}
\maketitle

\section{Introduction}

An easy way to explain the acceleration expansion of the Universe (see
\cite{Teg04,essence,Kowal08,komatsu08,BasPil}) is to consider an additional
fluid which has a negative equation of state parameter. This new fluid
counteracts the gravitational force and leads to the observed acceleration expansion.

Nevertheless, a plenitude of alternative cosmological scenarios has been a
result of the lack of a fundamental physical theory concerning the mechanism
that induces the cosmic acceleration. Most of them are based either on the
existence of new fields in nature (dark energy) or in some modification of the
Einstein-Hilbert action
\cite{BransDicke,Ratra88,NojiriS,Starobi,CapHyb,Horava,Pederico,Linder2004,YCai,Ycai1,Tamanini,Moraes,Bellido,Marco,Cremonini}%
.

One approach has been the consideration of two interacting scalar fields in a
spatially flat FRW spacetime . In this approach \ the interaction of the
scalar fields is usually limited to the potentials of the fields. This
limitation is not necessary and one would like to know what happens if the
interaction is extended to include as well the kinematics of the two fields.
In this case the dynamical system becomes quite more complicated and the
finding of analytic solutions is a major roblem. Indeed in the literature one
finds\ only a few successful attempts which find analytic solutions of the
field equations for the extended interaction. In \cite{arefeva} the authors
applied the superpotential method in order to determine a stable exact
solution. Another exact solution with two scalar fields with exponential
potential in which the one scalar field acts as a stiff matter given in
\cite{ChimentoTwoSF}. Finally in \cite{Moraes,Bazeia} it is shown that by
using the deformation procedure it is possible to generate a two scalar field
cosmological model from one scalar field, and use the solution of the single
scalar field to determine analytic solutions o the two scalar field model. We
note that in the case of a single scalar field there exists a large number of
analytic solutions (for instance see
\cite{MuslimovSF,MendezExactSF,EllisExactSF,HalliwellSF,Easther,Barrow,EPid,HKIM,CRoshan}%
).

The action of two interacting scalar fields in their kinematic and potential
part is \cite{Tamanini,Moraes,Bellido,Marco,Cremonini}
\begin{equation}
S=\int dx^{4}\sqrt{-g}\left(  R-\frac{1}{2}g_{ij}H_{AB}\left(  \Phi
^{C}\right)  \Phi^{A,i}\Phi^{B,i}+V\left(  \Phi^{C}\right)  \right)
\label{TF.01}%
\end{equation}
where $\Phi^{A}=\left(  \phi,\psi\right)  ~$and $H_{AB}=H_{AB}\left(  \Phi
^{C}\right)  =H_{BA}$ is a symmetric tensor. The importance of action
(\ref{TF.01}) is that a plethora of alternative theories of gravity can be
written in this form under a conformal transformation, for details see
\cite{AStaT}.

The main purpose of the present work is to address the problem of finding
analytic solutions of the two scalar field cosmology (\ref{TF.01}) in a
systematic way using the Noether symmetries of the field equations. As it will
be shown the method we propose recovers the aforementioned solutions and
produces, in addition, new ones which have not been considered.

The idea to apply Noether symmetries in scalar field cosmology and on modified
theories of gravity is not new and indeed it has gained a lot of attention in
the literature
\cite{YiZhang,deRiti90,Cap93deR,Cap97M,CapP09,CamL,KotsakisL,VakF,VF12,Sanyal05,Sanyal10,Kucuk13,Christodoulakis2013,dimakisT,deSouza,Hwei,FDarabi}%
. Not all approaches follow the same methods. In the present work we follow
the approach of \cite{Basilakos,TPBC,AnST,PalFR,BasFT,AndFT}\ which is
geometric hence, to our opinion, more fundamental.

The key to our approach is the recent result \cite{Tsam2d} that the Noether
point symmetries of Lagrange equations, for a first order Lagrangian, are the
homothetic vectors\footnote{We recall that a homothetic vector (HV) $X$ of a
metric $g_{ij}$ is a vector satisfying the identity $L_{X}g_{ij}=2\psi g_{ij}$
where $\psi$ is a constant. In case $\psi=0$ $\ $the vector $X$ is a Killing
vector (KV).} of the metric of the space generated by the dynamic fields. In
this way the determination of the Noether point symmetries becomes a problem
of differential geometry. Fortunately this problem nowadays can be dealt
easily with the use of appropriate software. For a general potential the field
equations do not admit Noether point symmetries hence they are not Noether
integrable. We demand that they admit extra Noether point symmetries which are
linearly independent and in involution, and determine in each case the
corresponding potentials using the results given in \cite{Tsam2d}.

Following the above we consider the symmetric tensor $H_{AB}$ in (\ref{TF.01})
as a metric in the space of the fields $\Phi^{C}$ and and apply the
aforementioned result to determine the Noether vectors and consequently the
corresponding Noether integrals which, provided that there are enough of them,
lead to the solution of the field equations. Obviously the approach depends
strongly on the metric $g_{ij}$ in (\ref{TF.01}). We show that in the flat
FRW\ background the requirement that the Lagrangian admits at least two
Noether point symmetries (apart form the trivial one $\partial_{t})\ $limits
the possible cases of interaction to two. One of them is the case considered
in \cite{AAslam} and the other, which is new, we consider and solve in detail
in section \ref{2dSO3}. In the latter case we find that the dynamical system
is equivalent to the motion of a particle in the $M^{3}$space. To find the
type of motion we use the Noether vectors to write the Lagrangian in normal
coordinates and in these coordinates it is found that the corresponding
potentials for which Noether symmetries are admitted, are equivalent to the 3d
unharmonic hyperbolic oscillator and to the 3d forced oscillator. Subsequently
in each case we determine easily the analytic solution.

The structure of the paper is as follows. In section \ref{FieldEquations} we
consider two scalar fields interacting both in their kinematic and potential
parts in a spatially flat FRW spacetime. We give the basic properties of the
dynamical system, we produce the field equations and we define the effective
equation of state. In section \ref{NoetherTheory} we discuss briefly the basic
theory of Noether point symmetries. We define the interaction of the scalar
fields in the kinetic part by the requirement that the field equations admit
at least two more Noether point symmetries. In section \ref{2dSO3} we
determine the potentials of the fields for which this is the case and then
solve analytically the resulting field equations. We find that for the first
potential the late time behavior of the scale factor is that of deSitter
solution. For the second potential the late time behavior is different. In
order to determine the late time behavior in this case we write the Hubble
function in terms of the redshifts and we find that that there exist a dust
like fluid component in the Hubble function.

Finally in order to show the viability of the new solution we compare it at
late time with the $\Lambda-$cosmology model using the supernova and the BAO
data. We find that both models fit the data practically with the same
statistic parameters. In section \ref{SONSF} we extend our analysis to the
case of $N$ scalar fields. In section \ref{ConEq} we show how the interaction
in the kinematic part of the scalar fields which we consider in section
\ref{2dSO3} arises from the conformal equivalence in scalar tensor theory.
Finally in section \ref{Conclusion} we draw our conclusions.

\section{The field equations}

\label{FieldEquations}

We assume that the fields interact in the spatially flat FRW spacetime
\begin{equation}
ds^{2}=-dt^{2}+a^{2}\left(  t\right)  \left(  dx^{2}+dy^{2}+dz^{2}\right)
\label{TF.01a}%
\end{equation}
where $a\left(  t\right)  $ is the scale factor. In this spacetime the
Lagrangian (\ref{TF.01}) becomes%

\begin{equation}
L\left(  a,\dot{a},\Phi^{C},\dot{\Phi}^{C}\right)  =-3a\dot{a}^{2}+\frac{1}%
{2}a^{3}H_{AB}\left(  \Phi^{C}\right)  \dot{\Phi}^{A}\dot{\Phi}^{B}%
-a^{3}V\left(  \Phi^{C}\right)  . \label{TF.02}%
\end{equation}
where the indices \thinspace$A,B,C$ $=1,2~$and $\Phi^{A}=\left(  \phi
,\psi\right)  .$ The field equations are the Friedmann equation:
\begin{equation}
-3a\dot{a}^{2}+\frac{1}{2}a^{3}H_{AB}\dot{\Phi}^{A}\dot{\Phi}^{B}%
+a^{3}V\left(  \Phi^{C}\right)  =0 \label{TF.03}%
\end{equation}
and the Euler Lagrange equations $\frac{\partial L}{\partial\left(  \dot
{a},\frac{d}{dt}\dot{\Phi}^{C}\right)  }-\frac{\partial L}{\partial\left(
a,\Phi^{C}\right)  }=0~$with respect to the variables $a,\Phi^{C}$; that is,
\begin{equation}
\ddot{a}+\frac{1}{2a}\dot{a}^{2}+\frac{a}{4}H_{AB}\dot{\Phi}^{A}\dot{\Phi}%
^{B}-\frac{1}{2}aV=0 \label{TF.03ab}%
\end{equation}%
\begin{equation}
\ddot{\Phi}^{A}+\frac{3}{2a}\dot{a}\dot{\Phi}^{A}+\tilde{\Gamma}_{BC}^{A}%
\dot{\Phi}^{B}\dot{\Phi}^{C}+H^{AB}V_{,B}=0 \label{TF.03a}%
\end{equation}
where $\tilde{\Gamma}_{BC}^{A}$ are the connection coefficients of the two
dimensional metric$~H_{AB}$.

In terms of the Hubble parameter $H=\frac{\dot{a}}{a}$ \ the field equations
(\ref{TF.03}) and (\ref{TF.03ab}) take the form%
\begin{equation}
H^{2}=\frac{1}{3}\left[  \frac{1}{2}H_{AB}\dot{\Phi}^{A}\dot{\Phi}%
^{B}+V\left(  \Phi^{C}\right)  \right]  \label{TF.03b}%
\end{equation}%
\begin{equation}
2\dot{H}+3H^{2}=-\left[  \frac{1}{2}H_{AB}\dot{\Phi}^{A}\dot{\Phi}%
^{B}-V\left(  \Phi^{C}\right)  \right]  . \label{TF.03c}%
\end{equation}

We assume commoving observers $u^{a}=\delta_{0}^{a}$ and from the field
equations we have the effective energy momentum tensor%
\begin{equation}
T_{ij}=\rho_{eff}u_{i}u_{j}+p_{eff}\left(  g_{ij}+u_{i}u_{j}\right)
\label{TF.03d}%
\end{equation}
where%
\begin{equation}
\rho_{eff}=\frac{1}{2}H_{AB}\dot{\Phi}^{A}\dot{\Phi}^{B}+V\left(  \Phi
^{C}\right)  \label{TF.03e}%
\end{equation}%
\begin{equation}
p_{eff}=\frac{1}{2}H_{AB}\dot{\Phi}^{A}\dot{\Phi}^{B}-V\left(  \Phi
^{C}\right)  . \label{TF.03f}%
\end{equation}
It follows that the "effective" equation of state is%
\begin{equation}
w_{eff}=\frac{\frac{1}{2}H_{AB}\dot{\Phi}^{A}\dot{\Phi}^{B}-V\left(  \Phi
^{C}\right)  }{\frac{1}{2}H_{AB}\dot{\Phi}^{A}\dot{\Phi}^{B}+V\left(  \Phi
^{C}\right)  }. \label{TF.03g}%
\end{equation}

Finally, the conservation equation $T_{;j}^{ij}=0$ gives
\begin{equation}
\dot{\rho}_{eff}+3\rho_{eff}H\left(  1+w_{eff}\right)  =0 \label{TF.03h}%
\end{equation}
from where we have the Klein Gordon equation (\ref{TF.03a}) for the fields.

As we have already mentioned for a general interaction (metric) $H_{AB}\left(
\Phi^{C}\right)  $ and potential$~V\left(  \Phi^{C}\right)  $ the dynamical
system defined by the Lagrangian (\ref{TF.02}) admits only the Noether point
symmetry $\partial_{t}$ with Noether integral the Hamiltonian (\ref{TF.03}).
In the following sections we apply the Noether symmetry approach in order to
determine both the kinematic interaction of the two fields - that is the
metric $H_{AB}\left(  \Phi^{C}\right)  $ - and the corresponding potential
$V\left(  \Phi^{C}\right)  $ for which two more Noether symmetries are
admitted, therefore the dynamical system is Liouville integrable.

\section{Noether point symmetries}

\label{NoetherTheory}

Before we proceed we review briefly the basic definitions concerning Noether
point symmetries of systems of second order ordinary differential equations
(ODEs) of the form%
\begin{equation}
\ddot{x}^{i}=\omega^{i}\left(  t,x^{j},\dot{x}^{j}\right)  . \label{Lie.0}%
\end{equation}

If the system of ODEs (\ref{Lie.0}) results from a first order Lagrangian,%
\begin{equation}
L=L\left(  t,x^{j},\dot{x}^{j}\right)  \label{Lie.1}%
\end{equation}
then the vector field
\[
X=\xi\left(  t,x^{j}\right)  \partial_{t}+\eta^{i}\left(  t,x\right)
\partial_{i}%
\]
in the augmented space $\{t,x^{i}\}$ is the generator of a Noether point
symmetry of the system (\ref{Lie.0}) if the following condition is
satisfied~\cite{StephaniB}
\begin{equation}
X^{\left[  1\right]  }L+L\frac{d\xi}{dt}=\frac{df}{dt} \label{Lie.5}%
\end{equation}
where $f=f\left(  t,x^{j}\right)  $\ is the Noether gauge function and
$X^{\left[  1\right]  }$ is the first prolongation \ of $X,$ i.e.%
\[
X^{\left[  1\right]  }=X+\left(  \frac{d\eta^{i}}{dt}-\dot{x}^{i}\frac{d\xi
}{dt}\right)  \partial_{\dot{x}^{i}}.
\]
To every Noether point symmetry there corresponds a first integral (a Noether
integral) of the system of equations (\ref{Lie.0}) given by the formula:%
\begin{equation}
I=\xi E_{H}-\frac{\partial L}{\partial\dot{x}^{i}}\eta^{i}+f \label{Lie.6}%
\end{equation}
where $E_{H}$ is the Hamiltonian of $L$.

The generator $X$ of a Noether point symmetry of the Lagrangian (\ref{TF.02})
in the space $\{t,a,\Phi^{C})$ is
\begin{equation}
X=\xi\left(  t,a,\Phi^{C}\right)  \partial_{t}+\eta_{a}\left(  t,a,\Phi
^{C}\right)  \partial_{a}+\eta_{\Phi}^{C}\left(  t,a,\Phi^{C}\right)
\partial_{\Phi^{C}} \label{Lie.3}%
\end{equation}
and the first prolongation vector is
\begin{equation}
X^{\left[  1\right]  }=X+\left(  \dot{\eta}_{a}-\dot{a}\dot{\xi}\right)
\partial_{\dot{a}}+\left(  \dot{\eta}_{\Phi}^{C}-\dot{\Phi}^{C}\dot{\xi
}\right)  \partial_{\dot{\Phi}^{C}}. \label{Lie.7}%
\end{equation}

Lagrangian (\ref{TF.02}) defines the kinetic metric%
\begin{equation}
ds_{\left(  3\right)  }^{2}=-6ada^{2}+a^{3}H_{AB}\left(  \Phi^{c}\right)
d\Phi^{A}d\Phi^{B} \label{TF.04}%
\end{equation}
and the effective potential $V_{eff}=a^{3}V\left(  \phi,\psi\right)  $.

It has been shown \cite{Tsam2d} that for Lagrangians of the form $T-V_{eff}$,
where $T$ is the kinetic energy, the Noether point symmetries are generated by
the elements of the homothetic group of the kinetic metric $T$. The metric
(\ref{TF.04}) can be written%
\begin{equation}
ds_{\left(  3\right)  }^{2}=a^{3}\left(  -\frac{6}{a^{2}}da^{2}+H_{AB}\left(
\Phi^{c}\right)  d\Phi^{A}d\Phi^{B}\right)  \label{TF.04a}%
\end{equation}
which shows that it is conformal to the 1+2 decomposable metric
\begin{equation}
d\tilde{s}_{\left(  3\right)  }^{2}=-\frac{6}{a^{2}}da^{2}+H_{AB}\left(
\Phi^{c}\right)  d\Phi^{A}d\Phi^{B}%
\end{equation}
where the 2d metric $H_{AB}$ is conformally flat (all 2d metrics are
conformally flat).

The 3d metric (\ref{TF.04}) for a general metric $H_{AB}\left(  \Phi
^{C}\right)  $ admits the gradient HV $H_{V}=\frac{2}{3}a\partial_{a}%
,~\psi_{H_{V}}=1~$\cite{Tupper89}. It is easy to show that the HV $H_{V}$ does
not generate a Noether point symmetry for the Lagrangian (\ref{TF.02}).

Therefore in order the Lagrangian (\ref{TF.02}) to admit additional Noether
point symmetries we have to consider special forms of the two dimensional
metric $H_{AB}\left(  \Phi^{C}\right)  $ for which the 3d metric (\ref{TF.04})
admits a greater homothetic algebra. Because we require two more first
integrals, apart from the Hamiltonian, the 2d metric\ $H_{AB}$ must be such
that the 3d metric (\ref{TF.04}) admits a homothetic algebra $\bar{G}_{HV}%
\,$\ with $\dim\bar{G}_{HV}\geq3.$ This happens in two cases:

Case 1: The 2d metric $H_{AB}$ admits three Killing vectors (KVs) which span
the $E\left(  2\right)  ~$group , i.e. $H_{AB}$ is flat.

Case 2: The 2d metric $H_{AB}$ admits three KVs which span the $SO\left(
3\right)  $ group, i.e. $H_{AB}$ is a space of constant non-vanishing curvature.

Before we proceed, we rewrite the Lagrangian (\ref{TF.02}) in a more
convenient form. We apply the coordinate transformation
\begin{equation}
a=\left(  \frac{3}{8}\right)  ^{\frac{1}{3}}u^{\frac{2}{3}} \label{TF}%
\end{equation}
and the Lagrangian (\ref{TF.02}) becomes%
\begin{equation}
L=-\frac{1}{2}\dot{u}+\frac{1}{2}u^{2}h_{AB}\dot{\Phi}^{A}\dot{\Phi}^{B}%
-u^{2}U\left(  \Phi^{C}\right)  \label{TF.05}%
\end{equation}
where we have set $h_{AB}=\frac{3}{8}H_{AB}$ and $U=\frac{3}{8}V.$~

In Case 1 the two dimensional metric $h_{AB}$ ~is $h_{AB}=\delta_{AB}.$ In the
case where $h_{AB}=\eta_{AB}$ the action (\ref{TF.01}) describes the Quintom
models \cite{ManosQuintom,TQiu}; this case has been considered in
\cite{AAslam}. Furthermore, this case is equivalent to that of a single
complex scalar field i.e.
\begin{equation}
S=\int dx^{4}\sqrt{-g}\left(  -\frac{R}{2}+\frac{1}{2}g_{ij}\Psi^{\ast i}%
\Psi^{i}+V\left(  \Psi,\Psi^{\ast}\right)  \right)
\end{equation}
where $\Psi=\phi+i\psi$ with product $\Psi\Psi^{\ast}=\left\vert
\Psi\right\vert ^{2}~$\cite{Fuzha1}. We remark that in the case where $h_{AB}$
is flat, we can always find a `coordinate' transformation in order to write
the kinetic term in the simplest form. For instance the mixed kinetic term\ in
\cite{ManosQuintom} can be reduced in this manner.

In Case 2 the 2d metric $h_{AB}$ is
\begin{equation}
h_{AB}=%
\begin{pmatrix}
1 & 0\\
0 & e^{2\phi}%
\end{pmatrix}
~ \label{Apen.00}%
\end{equation}
which is the metric of a maximally symmetric space of negative curvature. This
case is also equivalent to the one of a single complex scalar field $\Psi
=\phi+i\psi$ with the non Euclidian product
\begin{equation}
\Psi\Psi^{\ast}=\left(  1-\frac{1}{4}\left\vert \Psi\right\vert ^{2}\right)
^{-2}\left(  \left\vert \Psi\right\vert ^{2}\right)
\end{equation}
where $\left\vert \Psi\right\vert $ is the Euclidian norm. In the following we
discuss in detail the new Case 2.

\section{The metric $H_{AB}$ admits the $SO\left(  3\right)  $ Lie algebra}

\label{2dSO3}

As it has been remarked in this case $h_{AB}$ is the metric of a space of
constant non vanishing curvature. From (\ref{Apen.00}) the kinetic metric
(\ref{Apen.01}) becomes
\begin{equation}
ds_{\left(  3\right)  }^{2}=-du^{2}+u^{2}\left(  d\phi^{2}+e^{2\phi}~d\psi
^{2}\right)  .
\end{equation}
It is easy to show that this metric is flat, therefore admits a seven
dimensional homothetic Lie algebra consisting of the following vectors:

- Three gradient KVs (translations) $K^{I},$ $I=1,2,3:$%
\begin{align*}
K^{1}  &  =-\frac{1}{2}\left(  e^{\phi}\left(  1+\psi^{2}\right)  +e^{-\phi
}\right)  \partial_{u}\\
&  ~~+\frac{1}{2u}\left(  e^{\phi}\left(  1+\psi^{2}\right)  -e^{-\phi
}\right)  \partial_{\phi}+\frac{1}{u}\psi e^{-\phi}\partial_{\psi}%
\end{align*}%
\begin{align*}
K^{2}  &  =-\frac{1}{2}\left(  e^{\phi}\left(  1-\psi^{2}\right)  -e^{-\phi
}\right)  \partial_{u}\\
&  ~~+\frac{1}{2u}\left(  e^{\phi}\left(  1-\psi^{2}\right)  +e^{-\phi
}\right)  \partial_{\phi}-\frac{1}{u}\psi e^{-\phi}\partial_{\psi}%
\end{align*}%
\[
K^{3}=-\psi e^{\phi}\partial_{u}+\frac{1}{u}\psi e^{\phi}\partial_{\phi}%
+\frac{1}{u}e^{-\phi}\partial_{\psi}%
\]
where the corresponding gradient Killing functions $S_{\left(  I\right)  }$
are%
\[
S_{\left(  1\right)  }=\frac{1}{2}u\left(  e^{\phi}\left(  1+\psi^{2}\right)
+e^{-\phi}\right)
\]%
\[
S_{\left(  2\right)  }=\frac{1}{2}u\left(  e^{\phi}\left(  1-\psi^{2}\right)
-e^{-\phi}\right)
\]%
\[
S_{\left(  3\right)  }=u\psi e^{\phi}.
\]

-Three non-gradient KVs (the rotations) which span the $SO\left(  3\right)  $
algebra
\begin{align}
X_{12}  &  =\partial_{\psi}~,~\ X_{23}=\partial_{\phi}+\psi\partial_{\psi
}\label{e1.01}\\
X_{13}  &  =\psi\partial_{\phi}+\frac{1}{2}\left(  \psi^{2}-e^{2\phi}\right)
\partial_{\phi}. \label{e1.02}%
\end{align}

-The gradient HV $H_{V}=u\partial_{u}~,~~\psi_{H_{V}}=1.$

For this choice of $H_{AB}$ the Lagrangian (\ref{TF.05}) becomes%
\begin{equation}
L=-\frac{1}{2}\dot{u}+\frac{1}{2}u^{2}\left(  \dot{\phi}^{2}+e^{2\phi}%
~\dot{\psi}^{2}\right)  -u^{2}U\left(  \Phi^{C}\right)  . \label{Apen.01}%
\end{equation}
The field equations are the Hamiltonian
\begin{equation}
E=-\frac{1}{2}\dot{u}+\frac{1}{2}u^{2}\left(  \dot{\phi}^{2}+e^{2\phi}%
~\dot{\psi}^{2}\right)  +u^{2}U\left(  \Phi^{C}\right)  =0 \label{Apen.02}%
\end{equation}
and the Euler Lagrange equations
\begin{align}
\ddot{u}+u\dot{\phi}^{2}+ue^{2\phi}~\dot{\psi}^{2}-2uU  &  =0\label{Apen.03}\\
\ddot{\phi}+\frac{2}{u}\dot{u}\dot{\phi}-e^{2\phi}\dot{\psi}^{2}+U_{,\phi}  &
=0\label{Apen.04}\\
\ddot{\psi}+\frac{2}{u}\dot{u}\dot{\psi}+2\dot{\phi}\dot{\psi}+e^{-2\phi
}U_{,\psi}  &  =0. \label{Apen.05}%
\end{align}

In the following we demand the field equations to admit two extra Noether
point symmetries which are linearly independent and in involution, so that the
dynamical system is Liouville integrable.

Because the Lagrangian (\ref{Apen.01}) describes the motion of a particle in a
three dimensional flat space we need to know all potentials $U\left(  \Phi
^{C}\right)  $ for which the Lagrangian admits the extra two Noether point
symmetries. The answer to this problem has been given in \cite{Tsam3d}. Indeed
in \cite{Tsam3d} all 3d dynamical systems (that is potentials) have been
determined which admit extra Noether point symmetries. These potentials as
well as the corresponding Noether vectors and the subsequent Noether integrals
are given for each case in the form of tables. In the following we use these
tables to get directly the results we need in our problem.

From the tables of \cite{Tsam3d} we read that there are only two possible
cases that the Noether generators span the $so(3)$ algebra, One is the case of
the unharmonic oscillator and the other is the case of the forced oscillator.

\subsection{Case A: The unharmonic oscillator}

\label{unhoscillator}

In this case the potential $U\left(  \phi,\psi\right)  $ is%
\begin{equation}
u^{2}U\left(  \phi,\psi\right)  =\frac{\omega_{1}^{2}}{2}S_{\left(  1\right)
}^{2}-\frac{\omega_{2}^{2}}{2}S_{\left(  2\right)  }^{2}-\frac{\omega_{3}^{2}%
}{2}S_{\left(  3\right)  }^{2}%
\end{equation}
where $S_{\left(  I\right)  }^{2},$ $I=1,2,3$ are the gradient KV functions of
the flat space. The extra Noether point symmetries are (see Table 6 line 1
of~\cite{Tsam3d} with $p=0$)
\[
T_{1}\left(  t\right)  K^{1}~,~T_{2}\left(  t\right)  K^{2}~,T_{3}\left(
t\right)  K^{3}%
\]
where
\begin{equation}
T_{,tt}^{I}=\omega_{~J}^{I}T^{J}~,~\omega_{~J}^{I}=diag\left(  \left(
\omega_{1}\right)  ^{2},\left(  \omega_{2}\right)  ^{2},\left(  \omega
_{3}\right)  ^{3}\right)
\end{equation}
with gauge functions $f_{\left(  I\right)  }=T_{I,t}S_{\left(  I\right)  }$
and corresponding Noether integrals%
\begin{equation}
I_{C}^{I}=T_{I}\frac{d}{dt}S_{\left(  I\right)  }-T_{I,t}S_{\left(  I\right)
}.
\end{equation}
This dynamical system is the 3d `unharmonic oscillator' which is a well known
integrable system. We observe that when $\psi=\psi_{0}=const~$(that is
$\dot{\psi}=0$)$,$ the potential reduces to the well known UDM potential
\cite{BasilLukes,Bertacca}, i.e. the Lagrangian (\ref{Apen.01}) becomes
\begin{equation}
L=-\frac{1}{2}\dot{u}+\frac{1}{2}u^{2}\dot{\phi}^{2}-\frac{u^{2}}{8}\left(
\bar{\omega}_{1}^{2}e^{2\phi}+\bar{\omega}_{2}^{2}e^{-2\phi}+\bar{\omega}%
_{3}^{2}\right)  \label{UD1}%
\end{equation}
where $\bar{\omega}_{1},\bar{\omega}_{2},\bar{\omega}_{3}$ are%
\[
\bar{\omega}_{1}^{2}=\omega_{1}^{2}\left(  1+\psi_{0}^{2}\right)  ^{2}%
+\omega_{2}^{2}\left(  1-\psi_{0}^{2}\right)  +4\omega_{3}^{2}\psi_{0}^{2}%
\]%
\[
\bar{\omega}_{2}^{2}=\omega_{1}^{2}+\omega_{2}^{2}%
\]%
\[
\bar{\omega}_{3}^{2}=2\omega_{1}^{2}\left(  1+\psi_{0}^{2}\right)
-2\omega_{2}^{2}\left(  1-\psi_{0}^{2}\right)  .
\]
In \cite{Basilakos} it has been shown that the Lagrangian (\ref{UD1})
describes the two dimensional hyperbolic oscillator .

\subsubsection{Normal Coordinates}

Under the coordinate transformation%
\begin{align}
x  &  =\frac{1}{2}u\left(  e^{\phi}\left(  1+\psi^{2}\right)  +e^{-\phi
}\right)  ~~\label{tr.01}\\
~~y  &  =\frac{1}{2}u\left(  e^{\phi}\left(  1-\psi^{2}\right)  -e^{-\phi
}\right) \label{tr.02}\\
z  &  =u\psi e^{\phi} \label{tr.03}%
\end{align}
with inverse%
\begin{equation}
u^{2}=x^{2}-y^{2}-z^{2}~~,~~\phi=\ln\frac{x+y}{\sqrt{x^{2}-y^{2}-z^{2}}%
}~~,~\psi=\frac{z}{x+y}%
\end{equation}
the Lagrangian (\ref{Apen.01}) becomes
\[
L=-\frac{1}{2}\dot{x}^{2}+\frac{1}{2}\dot{y}+\frac{1}{2}\dot{z}^{2}%
-\frac{\omega_{1}^{2}}{2}x^{2}+\frac{\omega_{2}^{2}}{2}y^{2}+\frac{\omega
_{3}^{2}}{2}z^{2}.
\]
In these coordinates the field equations (\ref{Apen.02})-(\ref{Apen.05}) and
the constraint equation are reduced as follows
\begin{align}
\ddot{x}-\left(  \omega_{1}\right)  ^{2}x  &  =0\\
\ddot{y}-\left(  \omega_{2}\right)  ^{2}y  &  =0\\
\ddot{z}-\left(  \omega_{3}\right)  ^{2}z  &  =0
\end{align}%
\begin{equation}
0=-\frac{1}{2}\dot{x}^{2}+\frac{1}{2}\dot{y}+\frac{1}{2}\dot{z}^{2}%
+\frac{\omega_{1}^{2}}{2}x^{2}-\frac{\omega_{2}^{2}}{2}y^{2}-\frac{\omega
_{3}^{2}}{2}z^{2}.
\end{equation}
The analytic solution for the scale factor is found easily to be
\begin{equation}
a\left(  t\right)  =\left(  \frac{3}{8}\right)  ^{\frac{1}{3}}\left[
B_{IJ}X^{I}X^{J}\right]  ^{\frac{1}{3}}%
\end{equation}
where~ $B_{IJ}=diag\left(  1,-1,-1\right)  $ and
\begin{equation}
X^{I}=\omega_{~J}^{I}X^{J}~\text{, where }\omega_{~~J}^{I}=diag\left(  \left(
\omega_{1}\right)  ^{2},\left(  \omega_{2}\right)  ^{2},\left(  \omega
_{3}\right)  ^{2}\right)  \text{ ; }I,J=1,2,3
\end{equation}
For instance, if $\omega_{1}\omega_{2}\omega_{3}\neq0$ the analytic solution
is%
\begin{align*}
x\left(  t\right)   &  =x_{1}\sinh\left(  \omega_{1}t+x_{2}\right) \\
y\left(  t\right)   &  =y_{1}\sinh\left(  \omega_{2}t+y_{2}\right) \\
z\left(  t\right)   &  =z_{1}\sinh\left(  \omega_{3}t+z_{2}\right)
\end{align*}
with Hamiltonian constraint%
\begin{equation}
\left(  \omega_{1}x_{1}\right)  ^{2}-\left(  \omega_{2}y_{1}\right)
^{2}-\left(  \omega_{3}z_{1}\right)  ^{2}=0 \label{cc.01}%
\end{equation}
and the scale factor takes the following form%
\begin{equation}
a^{3}\left(  t\right)  =\frac{3}{8}\left[
\begin{array}
[c]{c}%
x_{1}^{2}\sinh^{2}\left(  \omega_{1}t+x_{2}\right)  +\\
-y_{1}^{2}\sinh^{2}\left(  \omega_{2}t+y_{2}\right)  -z_{1}^{2}\sinh
^{2}\left(  \omega_{3}t+z_{2}\right)
\end{array}
\right]  .
\end{equation}
We note that at late time the scale factor follows an exponential law, i.e.
$a\left(  t\right)  \varpropto e^{\omega t}$. Furthermore from the singularity
condition $a\left(  t\rightarrow0\right)  =0$ we have the additional
constraint equation%
\begin{equation}
x_{1}^{2}\sinh^{2}x_{2}-y_{1}^{2}\sinh^{2}y_{2}-z_{1}^{2}\sinh^{2}\left(
z_{2}\right)  =0. \label{cc.02}%
\end{equation}
Due to the constraints (\ref{cc.01}) and (\ref{cc.02}) the free parameters of
the model are four.

When $\omega_{\mu}=\omega_{\nu}$, $\ $i.e.
\[
\det\omega_{J}^{I}=\omega_{I}^{4}\omega_{J}^{2}~,~I\neq J~~;I,J=1,2,3
\]
the dynamical system admits one extra Noether symmetry. This is the rotation
normal to the plane defined by the $x^{I},x^{J}$ axes and it is generated by
the vector $X=x^{I}\partial_{J}-\varepsilon x^{J}\partial_{I}$ where
$\varepsilon=-1$ if $x^{I}/x^{J}=1$ and $\varepsilon=1$ if $x^{I}/x^{J}\neq1$.
$\ $Finally when~$\det\omega_{J}^{I}=\omega_{J}^{6},~$the potential $V\left(
\phi,\psi\right)  =V_{0}$ and the dynamical system is the 3d hyperbolic
oscillator (or the free particle if all $\omega_{I}=0)$ and admits 12 Noether
point symmetries (including the $\partial_{t})~~$\cite{Tsam3d,Wulfman}. That
means that the Noether point symmetries can also be used in order to reduce
the number of free parameters.

\subsection{Case B: The forced oscillator}

\label{CASE B}In this case the potential is$~$%
\begin{equation}
u^{2}U\left(  \phi,\psi\right)  =\frac{\omega_{0}^{2}}{2}u^{2}+\frac{\mu^{2}%
}{2\left(  1-a_{0}^{2}\right)  }\left(  S_{\left(  I\right)  }+a_{0}S_{\left(
J\right)  }\right)  ^{2}-\frac{\omega_{3}^{2}}{2}S_{\left(  K\right)  }^{2}
\label{CB.1}%
\end{equation}
where $a_{0}\neq1$; from Table A.1 line 1 of \cite{Tsam3d} we read that the
Noether point symmetries are
\begin{equation}
X_{1}=\bar{T}\left(  t\right)  \left(  K^{I}+a_{0}K^{J}\right)  ~~,~X_{2}%
=T_{K}\left(  t\right)  K^{K}~ \label{CB.2}%
\end{equation}%
\[
~X_{3}=T^{\ast}\left(  t\right)  \left(  a_{0}K^{I}+K^{J}\right)
\]
where the functions $\bar{T},T^{\ast},T_{I}$ are the solutions of the system%
\begin{equation}
\bar{T}_{,tt}=\left(  \mu^{2}+\omega_{0}^{2}\right)  \bar{T}~,~T_{I,tt}%
=\left(  \omega_{3}^{2}+\omega_{0}^{2}\right)  T_{I}~ \label{CB.3}%
\end{equation}%
\[
~T_{,tt}^{\ast}=\omega_{0}^{2}\bar{T}.
\]
The gauge functions are{\LARGE .}%

\[
f_{1}=\bar{T}_{,t}\left(  S_{\left(  \mu\right)  }+a_{0}S_{\left(  \nu\right)
}\right)  ,~f_{2}=T_{\sigma,t}S_{\left(  \sigma\right)  }%
\]
and
\[
f_{3}=T_{,t}^{\ast}\left(  a_{0}S_{\left(  \mu\right)  }+S_{\left(
\nu\right)  }\right)  .
\]
Hence the corresponding Noether integrals are%
\begin{equation}
\bar{I}_{1}=\bar{T}\frac{d}{dt}\left(  S_{\left(  I\right)  }+a_{0}S_{\left(
J\right)  }\right)  -\bar{T}_{,t}\left(  S_{\left(  I\right)  }+a_{0}%
S_{\left(  J\right)  }\right)  \label{CB.4}%
\end{equation}%
\begin{equation}
I_{2}=T_{\sigma}\frac{d}{dt}S_{\left(  K\right)  }-T_{\sigma,t}S_{\left(
K\right)  } \label{CB.5}%
\end{equation}%
\begin{equation}
I_{3}=T^{\ast}\frac{d}{dt}\left(  a_{0}S_{\left(  I\right)  }+S_{\left(
J\right)  }\right)  -T_{,t}^{\ast}\left(  a_{0}S_{\left(  I\right)
}+S_{\left(  J\right)  }\right)  .
\end{equation}

In order to continue we select $I=1~,~J=2,~K=3$.

\subsubsection{Normal Coordinates}

In case \ $I=1~,~J=2,~K=3$ the potential (\ref{CB.1}) becomes%
\begin{equation}
u^{2}U\left(  \phi,\psi\right)  =\frac{\omega_{0}^{2}}{2}u^{2}+\frac{\mu^{2}%
}{2\left(  1-a_{0}^{2}\right)  }\left(  S_{\left(  1\right)  }+a_{0}S_{\left(
2\right)  }\right)  ^{2}-\frac{\omega_{3}^{2}}{2}S_{\left(  3\right)  }^{2}
\label{CB.6}%
\end{equation}
where $,~a_{0}\neq1$. Under the coordinate transformation%
\begin{equation}
x=\left(  w+v\right)  ~,~~y=\frac{1}{a_{0}}\left(  w-v\right)  ~,~z=z
\label{tr.11}%
\end{equation}
where the variables $\left(  x,y,z\right)  $ follow from (\ref{tr.01}%
)-(\ref{tr.03}) the Lagrangian (\ref{Apen.01}) becomes%
\begin{equation}
L=T_{NC}-V_{NC} \label{CB.7}%
\end{equation}
where $T_{NC}$ is the kinetic energy in the coordinates $\left(  w,v,z\right)
$
\begin{equation}
T_{NC}=\frac{1}{2}\left[  \left(  \frac{1}{a_{0}^{2}}-1\right)  \dot{w}%
^{2}-\left(  \frac{1}{a_{0}^{2}}+1\right)  dwdv+\left(  \frac{1}{a_{0}^{2}%
}-1\right)  dv^{2}+\frac{1}{2}z^{2}\right]
\end{equation}
and $V_{NC}$ is the effective potential%
\begin{equation}
V_{NC}=-\frac{2\mu^{2}}{\left(  a_{0}^{2}-1\right)  }w^{2}-\frac{1}{2}\left(
\omega_{3}^{2}+\omega_{0}^{2}\right)  z^{2}+\frac{\omega_{0}^{2}}{2}\left(
\left(  w+v\right)  ^{2}-\frac{1}{a_{0}^{2}}\left(  w-v\right)  ^{2}\right)  .
\end{equation}
From this Lagrangian the field equations (\ref{Apen.02})-(\ref{Apen.05}) and
the Hamiltonian constraint become%
\begin{equation}
\ddot{w}-\left(  \mu^{2}+\omega_{0}^{2}\right)  w=0 \label{CB.8}%
\end{equation}%
\begin{equation}
\ddot{v}+\frac{a_{0}^{2}+1}{a_{0}^{2}-1}\mu^{2}w-\omega_{0}^{2}\nu=0
\label{CB.9}%
\end{equation}%
\begin{equation}
\ddot{z}-\left(  \omega_{3}^{2}+\omega_{0}^{2}\right)  z=0 \label{CB.10}%
\end{equation}%
\begin{equation}
T_{NC}+V_{NC}=0. \label{CB.11}%
\end{equation}

This system can be solved easily. For instance in the case where $\mu
^{2}\omega_{0}^{2}\neq0\,$\ the exact solution is%
\[
w\left(  t\right)  =w_{1}\exp\left(  \sqrt{\mu^{2}+\omega_{0}^{2}}t\right)
+w_{2}\exp\left(  -\sqrt{\mu^{2}+\omega_{0}^{2}}t\right)
\]%
\[
z\left(  t\right)  =z_{1}\exp\left(  \sqrt{\omega_{3}^{2}+\omega_{0}^{2}%
}t\right)  +z_{1}\exp\left(  -\sqrt{\omega_{3}^{2}+\omega_{0}^{2}}t\right)
\]%
\begin{equation}
v\left(  t\right)  =\frac{1+a_{0}^{2}}{1-a_{0}^{2}}w\left(  t\right)
+v_{1}\exp\omega_{0}t+v_{2}\exp\left(  -\omega_{0}t\right)
\end{equation}
with Hamiltonian constraint%
\[
-2\left(  \omega_{3}^{2}+\omega_{0}^{2}\right)  z_{1}z_{2}+\frac{8\left(
\mu^{2}+\omega_{0}^{2}\right)  }{1-a_{0}^{2}}w_{1}w_{2}+\frac{2}{a}\left(
a_{0}^{2}-1\right)  \omega_{0}^{2}v_{1}v_{2}=0.
\]
We note that for this potential the scale factor at late time follows also an
exponential law.

\subsubsection{Subcase B.1}

In this case the potential is
\begin{equation}
u^{2}U\left(  \phi,\psi\right)  =\frac{\omega_{0}^{2}}{2}u^{2}+\frac{\mu^{2}%
}{2}\left(  S_{\left(  I\right)  }+S_{\left(  J\right)  }\right)  ^{2}%
+\frac{\omega_{3}^{2}}{2}S_{\left(  K\right)  }^{2}. \label{sc.00}%
\end{equation}
The dynamical system admits the Noether point symmetries%
\begin{equation}
\bar{X}_{1}=\bar{T}\left(  t\right)  \left(  K^{I}+K^{J}\right)  ~~,~\bar
{X}_{2}=T_{K}\left(  t\right)  K^{K}.
\end{equation}
The functions $\bar{T}\left(  t\right)  ,T_{K}\left(  t\right)  $ follow from
(\ref{CB.2}) and the corresponding Noether integrals are given in
(\ref{CB.4}),(\ref{CB.5}).

In the case where $I=1,~J=2$ and $K=3$ under the coordinate transformation
(\ref{tr.11}) (for $a_{0}=1$) the Lagrangian (\ref{Apen.01}) becomes%
\[
L=-2\dot{w}\dot{v}+\frac{1}{2}\dot{z}^{2}-4\mu^{2}w^{2}-2\omega_{0}%
^{2}wv+\frac{1}{2}\left(  \omega_{3}^{2}+\omega_{0}^{2}\right)  z^{2}%
\]
and the field equations (\ref{Apen.02})-(\ref{Apen.05}) and the Hamiltonian
constraint become%
\begin{equation}
\ddot{w}-\omega_{0}^{2}w=0
\end{equation}%
\begin{equation}
\ddot{v}-4\mu^{2}w-\omega_{0}^{2}v=0
\end{equation}%
\begin{equation}
\ddot{z}-\left(  \omega_{3}^{2}+\omega_{0}^{2}\right)  z=0
\end{equation}%
\begin{equation}
0=-2\dot{w}\dot{v}+\frac{1}{2}\dot{z}^{2}+4\mu^{2}w^{2}+2\omega_{0}%
^{2}wv-\frac{1}{2}\left(  \omega_{3}^{2}+\omega_{0}^{2}\right)  z^{2}.
\end{equation}

The analytic solution of this system of equations is
\begin{equation}
w\left(  t\right)  =w_{1}\exp\left(  \omega_{0}t\right)  +w_{2}\exp\left(
-\omega_{0}t\right)  \label{sc.01}%
\end{equation}%
\begin{equation}
z\left(  t\right)  =z_{1}\exp\left(  \sqrt{\omega_{3}^{2}+\omega_{0}^{2}%
}t\right)  +z_{2}\exp\left(  -\sqrt{\omega_{3}^{2}+\omega_{0}^{2}}t\right)
\label{sc.02}%
\end{equation}%
\begin{align}
v\left(  t\right)   &  =\left(  2\omega_{0}t-v_{1}\right)  \mu^{2}\frac{w_{1}%
}{\omega_{0}^{2}}\exp\left(  \omega_{0}t\right)  +\label{sc.03}\\
&  ~~~-\left(  2\omega_{0}t+v_{2}\right)  \mu^{2}\frac{w_{2}}{\omega_{0}^{2}%
}\exp\left(  -\omega_{0}t\right)
\end{align}
with Hamiltonian constraint
\[
2\mu^{2}w_{1}w_{2}\left(  v_{1}+v_{2}-4\right)  +z_{1}z_{2}\left(  \omega
_{3}^{2}+\omega_{0}^{2}\right)  =0.
\]

As it is the case with case A for special values of the parameters $\omega
_{0},\mu,\omega_{3}$ it is possible that the dynamical system admits more
Noether point symmetries, which are produced by the elements of $SO\left(
3\right)  .$ However this adds nothing to the integrability of the system and
there is no point to consider these cases further. \ 

From (\ref{sc.01})-(\ref{sc.03}) and (\ref{tr.11}) we have that at late time
the scale factor has the following functional form
\begin{equation}
a\left(  t\right)  =\left(  3\frac{w_{1}^{2}}{\omega_{0}}\mu^{2}\right)
^{\frac{1}{3}}t^{\frac{1}{3}}e^{\frac{2}{3}\omega_{0}t}. \label{sf.01}%
\end{equation}

We can reconstruct this solution for the scale factor by selecting $\left(
w_{2},z_{2},\omega_{3}\right)  =0~$in (\ref{sc.01})-(\ref{sc.03}) and by
applying the singularity condition $a\left(  t\rightarrow0\right)  =0.$
Moreover for the scale factor (\ref{sf.01}) the Hubble function takes the form%
\begin{equation}
H\left(  t\right)  =\frac{2}{3}\omega_{0}+\frac{1}{3}t. \label{sf.02}%
\end{equation}
This solution is of interest because it is also the late time behavior of the
scale factor for values $\left(  w_{2},z_{2},\omega_{3}\right)  \neq0$.
However solution (\ref{sf.01}) identifies the second scalar field to be a
constant, i.e. $\psi\left(  t\right)  =\psi_{0}$.

In order to study the late time behaviour of (\ref{sf.01}) we write the Hubble
function (\ref{sf.02}) in terms of the redshift $z,~\frac{a_{0}}{a}=1+z$~where
$a_{0}$ is the renormalized parameter so that $a\left(  t_{today}\right)  =1$.
From equation (\ref{sf.01}) it follows that the parameter $t$ in terms of the
scale factor \ is expressed as follows%
\[
t=\frac{1}{2\omega_{0}}W\left(  \frac{2}{3}\frac{\omega_{0}^{2}}{w_{1}^{2}%
\mu^{2}}a^{3}\right)
\]
where $W\left(  x\right)  $ is the Lambert-W function. Then the Hubble
function (\ref{sf.02}) becomes%
\begin{equation}
H\left(  z\right)  =\frac{2}{3}\omega_{0}\left(  1+\left(  W\left(  \frac
{c}{\left(  1+z\right)  ^{3}}\right)  \right)  ^{-1}\right)  \label{sf.03}%
\end{equation}
where $c=\frac{2}{3}\frac{a_{0}\omega_{0}^{2}}{w_{1}^{2}\mu^{2}}$, $\omega
_{0}=\bar{\omega}H_{0}.~H_{0}$ is the Hubble constant for the present time
$H\left(  0\right)  =H_{0}$, hence from (\ref{sf.03}) we have the constrain
condition%
\[
c=\frac{2\bar{\omega}_{0}}{3-2\bar{\omega}_{0}}\exp\left(  \frac{2\bar{\omega
}_{0}}{3-2\bar{\omega}_{0}}\right)  .
\]

Finally the free parameters of the model are the $\bar{\omega}_{0}$ and the
Hubble constant $H_{0}$. We consider the Taylor expansion of the Hubble
function near the present time, i.e. $z=0$
\begin{equation}
H\left(  z\right)  =a_{0}+a_{1}z+a_{2}z^{2}+a_{3}z^{3}+O\left(  z^{4}\right)
\label{sf.04}%
\end{equation}
Using (\ref{sf.03}) we find that the constants $a_{0},a_{1},a_{2}$ and $a_{3}$
are%
\[
a_{0}=\frac{2}{3}\bar{\omega}_{0}H_{0}\left(  1+\frac{1}{W_{c}}\right)
~~,~a_{1}=\frac{2\bar{\omega}_{0}H_{0}}{\left(  1+W_{c}\right)  W_{c}}%
\]%
\[
a_{2}=\bar{\omega}_{0}H_{0}\frac{4W_{c}+2-W_{c}^{2}}{\left(  1+W_{c}\right)
^{3}W_{c}}%
\]%
\[
a_{3}=\bar{\omega}_{o}H_{0}\frac{\left(  2W_{c}^{4}-10W_{c}^{3}+21W_{c}%
^{2}+8W_{c}+2\right)  }{3\left(  1+W_{c}\right)  ^{5}W_{c}}%
\]
and $W_{c}=W\left(  c\right)  $. Therefore the Hubble function (\ref{sf.04})
can be written in the form%
\begin{equation}
H\left(  z\right)  =H_{0}\sqrt{\Omega_{m0}\left(  1+z\right)  ^{3}+\left(
1-\Omega_{m0}\right)  f_{DE}\left(  z,\bar{\omega}_{0}\right)  } \label{sf.05}%
\end{equation}
where we have defined $\Omega_{m0}=\frac{8}{9}\bar{\omega}_{0}^{2}%
\frac{\left(  W_{c}^{3}-6W_{c}^{2}+12W_{c}+10\right)  }{\left(  1+W_{c}%
\right)  ^{3}W_{c}^{2}}~$to be the density at the present time of the dust
like fluid (dark matter) and the function $f_{DE}\left(  z,\bar{\omega}%
_{0}\right)  $ describes the effective dark energy fluid.

\subsubsection{Cosmological Constrains of the late time solution}

We proceed with a joint likelihood analysis of the cosmological solution
(\ref{sf.01}) by using the Type Ia supernova data set of Union 2.1
\cite{Suzuki} and the 6dF, the SDSS and WiggleZ BAO data
\cite{Percival,BlakeC}. The likelihood function is defined as follows%
\begin{equation}
\mathcal{L}\left(  \mathbf{p}\right)  \mathcal{=L}_{SNIa}\mathcal{\times
L}_{BAO}%
\end{equation}
where $\mathbf{p}$ are the constrain parameters and $\mathcal{L}_{A}\varpropto
e^{-\chi_{A}^{2}/2}~$; that is, $\chi^{2}=\chi_{SNIa}^{2}+\chi_{BAO}^{2}$.
\ The Union 2.1 data set provide us with 580 SNIa distance modulus at observed
redshift. The chi-square parameter is given by the expression\footnote{For the
SNIa test we have applied the diagonal covariant matrix without the systematic
errors}%
\begin{equation}
\chi_{SNIa}^{2}=\sum\limits_{i=1}^{N_{SNIa}}\left(  \frac{\mu_{obs}\left(
z_{i}\right)  -\mu_{th}\left(  z_{i};\mathbf{p}\right)  }{\sigma_{i}}\right)
^{2}%
\end{equation}
where $z_{i}$ is the observed redshift ~in the range $0.015\leq z_{i}%
\leq1.414$~and $\mu$ is the distance modulus
\begin{equation}
\mu=m-M=5\log D_{L}+25
\end{equation}
and $D_{L}$ is the luminosity distance. For the constraint with the BAO data
the corresponding chi-square parameter is defined as follows%
\begin{equation}
\chi_{BAO}^{2}=\sum\limits_{i=1}^{N_{BAO}}\left(  \sum\limits_{j=1}^{N_{BAO}%
}\left[  d_{obs}\left(  z_{i}\right)  -d_{th}\left(  z_{i};\mathbf{p}\right)
\right]  C_{ij}^{-1}\left[  d_{obs}\left(  z_{j}\right)  -d_{th}\left(
z_{j};\mathbf{p}\right)  \right]  \right)
\end{equation}
where $N_{BAO}=6$, $C_{ij}^{-1}$ is the inverse of the covariant matrix in
terms of $d_{z}~\ $(see \cite{BasilNess}), and the parameter $d_{z}$ follows
from the relation $d_{z}=\frac{l_{BAO}}{D_{V}\left(  z\right)  }$;
$~l_{BAO}\left(  z_{drag}\right)  $ is the BAO scale at the drag redshift and
$D_{V}\left(  z\right)  $ is the volume distance \cite{BlakeC}.

For the Hubble constant we consider the value $H_{0}=69.6~km~s^{-1}Mpc^{-1}$
(see \cite{BennetH0}) hence the free parameter of the Hubble function
(\ref{sf.05}) is $\mathbf{p}=\Omega_{m0}$. \ Therefore, we find that the best
fit value~$\bar{\omega}_{m0}$ of the model (\ref{sf.05}) we derived is
$\Omega_{m0}=0.31_{-0.024}^{+0.023}$, with $\min\chi_{total}^{2}=564.8$; the
corresponding value of the constant $\bar{\omega}_{0}~$is $\bar{\omega}%
_{0}=0.925$. When $f_{DE}\left(  z,\bar{\omega}_{0}\right)  =1$ in
(\ref{sf.05}) the Hubble function (\ref{sf.05}) reduces to that of the
$\Lambda-$cosmology. Therefore, by constraining the $\Lambda-$cosmology with
the SNIa and the BAO data we find the minimum chi-square parameter $\min
{}_{\Lambda}\chi_{total}^{2}=564.5$ with matter density $\Omega_{m0}%
=0.28$.$_{-0.015}^{+0.017}$. We note that the difference between the minimum
chi-square parameters is $\min\left(  \chi_{total}^{2}-{}_{\Lambda}%
\chi_{total}^{2}\right)  =0.3$ which leads to the conclusion that both the
model (\ref{sf.05}) and the $\Lambda-$cosmology model fit the SNIa and the BAO
data with similar statistic parameters. This proves theviabiliy of the
solution we have found.

\section{The case of $N$ interacting scalar fields}

\label{SONSF}

As we have seen in section \ref{unhoscillator} the UDM model
\cite{BasilLukes,Bertacca} at the level of Noether symmetries (but also as a
dynamical system) is equivalent to the unharmonic oscillator \cite{Basilakos}.
In the case of two scalar fields this happens if the fields interact in their
kinematic part so that 2d the interaction metric $H_{AB}$ admits $so(3)$ as
the Killing algebra. In this section, we consider the case of $N$ scalar
fields which interact in their kinematic and potential parts with action
\cite{Berglund}
\begin{equation}
S=\int dx^{4}\sqrt{-g}\left(  R-\frac{1}{2}g_{ij}G_{a\beta}\Phi^{a,i}%
\Phi^{\beta,i}+V\left(  \Phi^{\zeta}\right)  \right)  \label{NSF.00}%
\end{equation}
where $G_{a\beta}$ is a second order symmetric tensor and $a,\beta=1,2...,N.$
We consider again $G_{a\beta}$ as a metric in the space spanned by the $N$
scalar fields. If we assume that $G_{a\beta}$ admits the algebra $so(N+1)\ $as
KVs, then $G_{ab}$ is the metric of a space of constant curvature and the
fundamental length $ds_{G}^{2}=G_{ab}d\Phi^{a}d\Phi^{\beta}~$can be written as
follows%
\begin{equation}
ds_{G}^{2}=d\Phi_{1}^{2}+e^{2\Phi_{1}}\left[  d\Phi_{2}^{2}+d\Phi_{3}%
^{2}+...+d\Phi_{N-1}^{2}\right]  . \label{NSF.01}%
\end{equation}
Assuming that the interaction takes place in the FRW spatially flat spacetime
(\ref{TF.01a}) the Lagrangian (\ref{NSF.00}) is
\begin{equation}
L\left(  a,\dot{a},\Phi^{\gamma},\dot{\Phi}^{\gamma}\right)  =-3a\dot{a}%
^{2}+\frac{1}{2}a^{3}G_{\gamma\delta}\dot{\Phi}^{\gamma,i}\dot{\Phi}%
^{\delta,i}-a^{3}V\left(  \Phi^{\zeta}\right)  . \label{NSF.02}%
\end{equation}

We introduce the new variable $u$ (see (\ref{TF})) and the Lagrangian
(\ref{NSF.02}) becomes
\[
L\left(  u,\dot{u},\Phi^{\gamma},\Phi^{\gamma}\right)  =-\frac{1}{2}\dot
{u}^{2}+\frac{1}{2}u^{2}G_{\gamma\delta}\dot{\Phi}^{\gamma,i}\dot{\Phi
}^{\delta,i}-u^{2}V\left(  \Phi^{\zeta}\right)
\]
in which we have introduced the effective potential
\begin{equation}
u^{2}V\left(  \Phi\right)  =\frac{1}{2}A_{IJ}S^{I}S^{J}~,~J=1..N+1
\label{NSF.02a}%
\end{equation}
where $A_{IJ}=-diag\left(  \left(  \omega_{1}\right)  ^{2},\left(  \omega
_{2}\right)  ^{2},...,\left(  \omega_{N}\right)  ^{2},-\left(  \omega
_{N+1}\right)  ^{2}\right)  $ and
\begin{align*}
S_{N+1}  &  =\frac{1}{2}u\left(  e^{\Phi_{1}}\left(  1+\Phi_{2}^{2}+\Phi
_{3}^{2}+...+\Phi_{N-1}^{2}\right)  +e^{-\Phi_{1}}\right) \\
S_{1}  &  =\frac{1}{2}u\left(  e^{\Phi_{1}}\left(  1-\left(  \Phi_{2}^{2}%
+\Phi_{3}^{2}+...+\Phi_{N-1}^{2}\right)  \right)  -e^{-\Phi_{1}}\right) \\
S_{2}  &  =ue^{\Phi}\Phi_{2}\\
S_{3}  &  =ue^{\Phi_{1}}\Phi_{3}\\
&  ...\\
S_{N}  &  =ue^{\Phi_{1}}\Phi_{N}.
\end{align*}
where $S_{N+1}$ are the gradient KVs of the flat space. Under the coordinate
transformation~$Z_{J}=S_{J}~$ the Lagrangian (\ref{NSF.02}) becomes
\begin{equation}
L\left(  Z^{J},\dot{Z}^{J}\right)  =-\frac{1}{2}\eta_{IJ}\dot{Z}^{I}\dot
{Z}^{J}-\frac{1}{2}A_{IJ}Z^{I}Z^{J}%
\end{equation}
where $\eta_{IJ}=diag\left(  1,1,...,1,-1\right)  $. Therefore the exact
solution of the scale factor is
\begin{equation}
a\left(  t\right)  =\left(  \frac{3}{8}\right)  ^{\frac{1}{3}}\left(
\eta_{IJ}Z^{I}Z^{J}\right)  ^{\frac{1}{3}}%
\end{equation}
where $Z^{I}\left(  t\right)  $ satisfies
\begin{equation}
\ddot{Z}^{I}-\eta^{IJ}A_{JK}Z^{K}=0 \label{NSF.04}%
\end{equation}
and
\begin{equation}
\frac{1}{2}\eta_{IJ}\dot{Z}^{I}\dot{Z}^{J}-\frac{1}{2}A_{IJ}Z^{I}Z^{J}=0.
\end{equation}

Note that equations (\ref{NSF.04}) describe the $\left(  N+1\right)  $
anisotropic oscillator. Another attempt to apply the Noether point symmetries
in the case of $N$ scalar fields can be found in \cite{YiZhang}. However in
\cite{YiZhang} the authors consider that the metric $G_{\alpha\beta}$ is
invariant under the $E\left(  N\right)  $ Lie algebra; that is, $G_{\alpha
\beta}$ is a flat space, and the scalar fields have the same potentials and
the same initial conditions, therefore the problem reduces to the one scalar
field cosmology.

\section{Conformal equivalence}

\label{ConEq}

In this section we study the interaction of two scalar fields under a
conformal transformation and we show how the cases of section \ref{2dSO3}
follow from the conformal equivalence in scalar tensor theory.

We assume that we have a model consisting of one non-minimally coupled scalar
field $\Phi$ and one minimally coupled scalar field $\psi.$ Then the action is%
\begin{equation}
S=\int dx^{4}\sqrt{-\bar{g}}\left[  F\left(  \Phi\right)  \bar{R}+\frac{1}%
{2}\bar{g}_{ij}^{;i}\Phi^{;i}\Phi^{;j}-\frac{1}{2}\bar{g}_{ij}\psi^{;i}%
\psi^{;j}-\bar{V}\left(  \Phi,\psi\right)  \right]  . \label{CT.0A}%
\end{equation}

Under the conformal transformation $\bar{g}_{ij}=N^{2}g_{ij}$ where
$N=\frac{1}{\sqrt{-2F\left(  \Phi\right)  }}$ the action becomes (see
\cite{TPBC,SCapdeRitis,Starobinsky1})
\begin{equation}
S=\int dx^{4}\sqrt{-g}\left[  -\frac{R}{2}+\frac{1}{2}\left(  \frac{3F_{\Phi
}^{2}-F}{2F^{2}}\right)  g_{ij}\Phi^{;i}\Phi^{;j}+\frac{1}{2}\frac
{1}{4F\left(  \Phi\right)  }g_{ij}\psi^{;i}\psi^{;j}-V\left(  \Phi
,\psi\right)  \right]
\end{equation}
where we have set%
\begin{equation}
V\left(  \Phi,\psi\right)  =\frac{\bar{V}\left(  \Phi,\psi\right)  }{4F^{2}}.
\end{equation}
If we consider the transformation $\Phi\rightarrow\phi$ by the formula%
\begin{equation}
d\phi=\sqrt{\left(  \frac{3F_{\Phi}^{2}-F}{2F^{2}}\right)  }d\Phi\label{CT.01}%
\end{equation}
the action $S$ takes the form:%
\begin{equation}
S=\int dx^{4}\sqrt{-g}\left[  -\frac{R}{2}+\frac{1}{2}g_{ij}\phi^{;i}\phi
^{;j}+\frac{1}{8}\frac{1}{F\left(  \phi\right)  }g_{ij}\psi^{;i}\psi
^{;j}-V\left(  \phi,\psi\right)  \right]
\end{equation}
which shows that the two scalar fields $\phi,\psi$ interact in their kinematic
part with the metric $g_{ij}=-2F\left(  \Phi\right)  \bar{g}_{ij}.$

To give an example of the above equivalence let us consider the simple case
where $F\left(  \Phi\right)  =f_{0}\Phi^{2},~f_{0}\neq\frac{1}{12}~$(see
\cite{SCapdeRitis}).Then from (\ref{CT.01}) we find $~\phi=C\ln\Phi$ where
$C=\frac{\sqrt{12F_{0}-1}}{\sqrt{2F_{0}}}$. Replacing we find that the action
becomes
\begin{equation}
S=\int dx^{4}\sqrt{-g}\left[  -\frac{R}{2}+\frac{1}{2}g_{ij}\phi^{;i}\phi
^{;j}+\frac{1}{8}e^{-C\phi}g_{ij}\psi^{;i}\psi^{;j}-V\left(  \phi,\psi\right)
\right]  \label{CT.02}%
\end{equation}
which implies that the two scalar fields $\phi,\psi$ interact with the two
dimensional metric \cite{Starobinsky1}%
\begin{equation}
ds_{\left(  2\right)  }^{2}=d\phi^{2}+\frac{1}{4}e^{-C\phi}d\psi^{2}.
\end{equation}
In the case where $C=2$, i.e. $f_{0}=-\frac{1}{6}$, the $ds_{\left(  2\right)
}^{2}$ is a space of constant curvature, and then the action (\ref{CT.02}) is
the one we considered in section \ref{2dSO3}.

We note that by replacing $F\left(  \Phi\right)  =-\frac{1}{6}\Phi^{2}$ and
$\Phi=\sqrt{6\zeta}$ in (\ref{CT.0A}) the action becomes%
\begin{equation}
S=-\int dx^{4}\sqrt{-\bar{g}}\left[  \zeta\bar{R}-\frac{3}{\zeta}\bar{g}%
_{ij}^{;i}\zeta^{;i}\zeta^{;j}+\frac{1}{2}\bar{g}_{ij}\psi^{;i}\psi^{;j}%
+\bar{V}\left(  \zeta,\psi\right)  \right]  .
\end{equation}
and $\zeta$ is a Brans-Dicke scalar field.

\section{Conclusion}

\label{Conclusion}

We have considered two scalar fields interacting both in their kinematic and
potential parts in a spatially flat FRW spacetime and determined those
interactions for which the dynamical system of the two scalar fields is
Liouville integrable. The system has three variables therefore for this to be
the case two more first integrals are required (in addition to the
Hamiltonian). One systematic way to find these integrals is to assure that the
Lagrangian admits another two Noether symmetries. According to recent results
\cite{Tsam2d,Tsam3d} this is possible if one uses the results relating the
Noether algebra with the homothetic algebra of the interaction metric $H_{AB}$
characterizing the interaction in the space of the fields. A\ detailed
examination of the Lagrangian shows that there are two cases to be considered
(1) the 2d metric $H_{AB}$ admits three KVs which span the $E\left(  2\right)
~$ group, i.e. $H_{AB}$ is flat. and (2) the 2d metric $H_{AB}$ admits three
KVs which span the $SO\left(  3\right)  $ group, i.e. $H_{AB}$ is a space of
constant non-vanishing curvature. The first case has been considered in
\cite{AAslam}. In the present work we considered the remaining case and showed
that in this case the dynamical systems which arise from the Noether
symmetries are (a) the unharmonic oscillator and (b) the forced oscillator. We
recall that the Lagrangian of the $\Lambda$- cosmology admits five Noether
symmetries and the resulting dynamical system is equivalent to the 1d
(hyperbolic) oscillator. In each case we determined the Noether symmetries and
the corresponding Noether integrals. Furthermore using the Noether vectors we
determined the normal coordinates and subsequently we solved analytically the
field equations.

In order to examine the viability of the solution we considered the late time
behavior of the scale factor and found that the scalar field introduces a dust
like component in the Hubble function. We performed a joint likelihood
analysis to constrain the model with the Supernova data of Union 2.1 and the
6dF, SDSS and WiggleZ BAO data, and we found that the model fits the
cosmological data with a minimum$~\chi_{total}^{2}=564.8$ and today's value of
the dark energy density $\Omega_{m0}=0.31_{-0.024}^{+0.023}.$ Comparing these
with the corresponding values of the $\Lambda-$cosmology model we find that
the difference between the two statistical parameters $\chi_{total}^{2}$of the
two models is $\Delta\chi_{total}^{2}=0.3$. This implies that the analytic
solution we have obtained mimics the $\Lambda-$cosmology at late time.

We generalized these considerations to the case of $N$ scalar fields
interacting both in their potential as well as in their kinematic part in a
flat FRW background and we computed again the analytic form of the scale
factor. Finally, we have shown that this type of interaction also follows from
a conformal transformation in the Brans-Dicke action.

Concluding we remark that it is possible to extend the symmetry method for the
action (\ref{TF.01}) in order to determine invariant solutions of the
Wheeler-DeWitt equation in quantum cosmology. Such an analysis is in progress
and it will be published in a forthcoming paper.

\begin{acknowledgments}
AP acknowledge financial support of INFN (initiative specifiche QGSKY, QNP,
and TEONGRAV)
\end{acknowledgments}

\appendix

\section{Representations of the $so\left(  3\right)  $ Lie algebra}

\label{ApRepr}

The form of the metric $h_{AB}$ (\ref{Apen.00}) we considered \ in section
\ref{2dSO3} corresponds to the representation of the $so\left(  3\right)  $
Lie algebra with elements (\ref{e1.01}),(\ref{e1.02}).

However it is possible to consider a different representation of the
$so\left(  3\right)  $ Lie algebra, therefore another form of the metric
$h_{AB}$. For instance another form of$~h_{AB}$ is%
\begin{equation}
d\bar{s}^{2}=d\bar{\phi}^{2}+\sinh^{2}\bar{\phi}d\bar{\psi}^{2}.
\label{TF.06a}%
\end{equation}
Obviously the two representations are related with a coordinate transformation
$\left(  \phi,\psi\right)  \rightarrow\left(  \bar{\phi},\bar{\psi}\right)  $
therefore our results remain true for the new representation. In order to show
this we consider the Lagrangian in which $h_{AB}$ is of the form
(\ref{TF.06a}). The Lagrangian is:%

\begin{equation}
L=-\frac{1}{2}\dot{u}+\frac{1}{2}u^{2}\left(  \left(  \overset{\cdot}%
{\bar{\phi}}\right)  ^{2}+\sinh^{2}\bar{\phi}\left(  \overset{\cdot}{\bar
{\psi}}\right)  ^{2}\right)  -u^{2}U\left(  \bar{\Phi}^{C}\right)  .
\label{TF.06}%
\end{equation}
For this Lagrangian the field equations and the Klein Gordon equations are
\begin{equation}
-\frac{1}{2}\dot{u}+\frac{1}{2}u^{2}\left(  \left(  \overset{\cdot}{\bar{\phi
}}\right)  ^{2}+\sinh^{2}\phi~\left(  \overset{\cdot}{\bar{\phi}}\right)
^{2}\right)  -u^{2}U\left(  \bar{\Phi}^{C}\right)  =0 \label{ATF.0}%
\end{equation}%
\begin{align}
\ddot{u}+u\left(  \overset{\cdot}{\bar{\phi}}\right)  ^{2}+u\sinh^{2}\bar
{\phi}~\left(  \overset{\cdot}{\bar{\phi}}\right)  ^{2}-2uU  &
=0\label{ATF.01A}\\
\overset{\cdot\cdot}{\bar{\phi}}+\frac{2}{u}\dot{u}\left(  \overset{\cdot
}{\bar{\phi}}\right)  -\sinh\bar{\phi}\cosh\bar{\phi}~\left(  \overset{\cdot
}{\bar{\psi}}\right)  ^{2}+U_{,\bar{\phi}}  &  =0\label{ATF.01B}\\
\overset{\cdot\cdot}{\bar{\psi}}+\frac{2}{u}\dot{u}\left(  \overset{\cdot
}{\bar{\psi}}\right)  +2\coth\bar{\phi}~\left(  \overset{\cdot}{\bar{\phi}%
}\right)  \left(  \overset{\cdot}{\bar{\psi}}\right)  +\sinh^{-2}\bar{\phi
}1U_{,\bar{\psi}}  &  =0. \label{ATF.01C}%
\end{align}

The kinetic metric of (\ref{TF.06}) is
\[
ds_{\left(  3\right)  }^{2}=-du^{2}+u^{2}\left[  d\bar{\phi}^{2}+\sinh^{2}%
\bar{\phi}~d\bar{\psi}^{2}\right]
\]
and the corresponding gradient KVs are%

\begin{align*}
\bar{S}_{\left(  1\right)  }  &  =u\cos\bar{\psi}\sinh\bar{\phi}\\
\bar{S}_{\left(  2\right)  }  &  =u\sin\bar{\psi}\sinh\bar{\phi}\\
\bar{S}_{\left(  3\right)  }  &  =u\cosh\bar{\phi}.
\end{align*}
By replacing the functions $\bar{S}_{\left(  1-3\right)  }$ instead of
$S_{\left(  1-3\right)  }$ in the potentials of section \ref{2dSO3} we find
the same exact solutions for the scale factor. Working similarly we show that
the result holds for the case of the $N$ scalar fields of section \ref{SONSF}.

\end{document}